\documentclass[3p,times]{elsarticle}

\usepackage{ecrc}

\volume{00}

\firstpage{1}

\runauth{W. Yu et al.}

\CopyrightLine{2023}{Published by Elsevier Ltd.}

\usepackage{framed,multirow}

\usepackage{amsmath,amssymb,amsfonts}
\usepackage{latexsym}

\usepackage{lineno}

\usepackage[figuresright]{rotating}

\newcommand{\methodname}{SAN-Net\xspace}

\newcommand{\mat}[1]{\boldsymbol{#1}}
\newcommand{\ie}{\emph{i.e.}\xspace}
\newcommand{\eg}{\emph{e.g.}\xspace}

\usepackage{url}
\usepackage{booktabs}
\usepackage{algorithmic}
\usepackage{graphicx}
\usepackage{textcomp}
\usepackage{multirow}
\usepackage{color}
\usepackage{booktabs} 
\usepackage{supertabular}
\usepackage{xspace}
\usepackage{makecell}
\usepackage{wrapfig}
\usepackage{hyperref}

\newlength\savewidth\newcommand\shline{\noalign{\global\savewidth\arrayrulewidth
  \global\arrayrulewidth 1.5pt}\hline\noalign{\global\arrayrulewidth\savewidth}}
\usepackage[pdftex,dvipsnames]{xcolor}

\begin{document}

\begin{frontmatter}

\dochead{}

\title{\methodname: Learning Generalization to  Unseen Sites for Stroke Lesion Segmentation with Self-Adaptive Normalization}

\author[1]{Weiyi Yu}
\author[2]{Zhizhong Huang}
\author[2]{Junping Zhang}
\author[1,3]{Hongming Shan\corref{cor1}}

\address[1]{Institute of Science and Technology for Brain-inspired
Intelligence and MOE Frontiers Center for Brain Science, \\Fudan University,
Shanghai, 200433, China}
\address[2]{Shanghai Key Lab of Intelligent Information Processing and the School of Computer Science,\\ Fudan University, Shanghai 200433, China}
\address[3]{Shanghai Center for Brain Science and Brain-inspired Technology, Shanghai 201210, China}
\cortext[cor1]{Corresponding author: hmshan@fudan.edu.cn}

\begin{abstract}
There are considerable interests in automatic stroke lesion segmentation on magnetic resonance (MR) images in the medical imaging field, as stroke is an important cerebrovascular disease. Although deep learning-based models have been proposed for this task, generalizing these models to unseen sites is difficult due to not only the large inter-site discrepancy among different scanners, imaging protocols, and populations, but also the variations in stroke lesion shape, size, and location. To tackle this issue, we introduce a self-adaptive normalization network, termed \methodname, to achieve adaptive generalization on unseen sites for stroke lesion segmentation. Motivated by traditional z-score normalization and dynamic network, we devise a masked adaptive instance normalization (MAIN) to minimize inter-site discrepancies, which standardizes input MR images from different sites into a site-unrelated style by dynamically learning affine parameters from the input; \ie, MAIN can affinely transform the intensity values. Then, we leverage a gradient reversal layer to force the U-net encoder to learn site-invariant representation with a site classifier, which further improves the model generalization in conjunction with MAIN. Finally, inspired by the ``pseudosymmetry'' of the human brain, we introduce a simple yet effective data augmentation technique, termed symmetry-inspired data augmentation (SIDA), that can be embedded within \methodname to double the sample size while halving memory consumption. 
Experimental results on the benchmark Anatomical Tracings of Lesions After Stroke (ATLAS) v1.2 dataset, which includes MR images from 9 different sites, demonstrate that under the ``leave-one-site-out'' setting, the proposed \methodname outperforms recently published methods in terms of quantitative metrics and qualitative comparisons.

\end{abstract}


\end{frontmatter}



\section{Introduction}\label{sec:intro}

As one of the common cerebrovascular diseases, stroke is the main cause of adult disability and the second leading cause of death worldwide~\cite{2020Heart}. Neuroimaging is commonly used to measure brain structures, aiding radiologists in understanding and predicting post-stroke brain changes~\cite{2009Interrater}. T1-weighted magnetic resonance imaging (MRI), a noninvasive neuroimaging technology, is widely used to examine structural changes in the patient's brain after a stroke. Manual delineation of stroke lesions is time-consuming and cumbersome; different radiologists may delineate various boundaries for the same lesion, and the same radiologist may not be able to reproduce their previous segmentation. Thus, automatic stroke lesion segmentation has become an important research topic over the past decade.

Currently, deep learning-based approaches,  especially convolutional neural networks (CNNs)~\cite{karthik2021delineation, karthik2022contour}, have made remarkable progress in medical image segmentation.
Considering that shape, size, and location of lesions vary substantially, training a precise segmentation model is challenging, especially when there are a limited number of training images in clinical practice. 
Many efforts have been made to address the problem of stroke lesion variability so far. In terms of 2D CNNs, the attention or multi-scale modules were designed to extract more long-range relationships~\cite{2019X, 2019MSDF}. In contrast, the fusion of 2D and 3D CNNs was a tradeoff between the number of model parameters and feature extraction capability~\cite{2019D, 2019A, DBLP:journals/sncs/BasakHR21}.

The existing stroke lesion segmentation methods were established assuming that MR images from different sites have consistent distributions. Thus, their generalization to unseen sites is challenging. Differences among MR scanners, imaging protocols, and variations in patient populations, greatly affect model generalization to unseen sites.
Therefore, many researchers pay attention to multi-site learning, such as domain adaptation and domain generalization.  Domain adaptation intends to narrow the domain gap between source and target domains, and there are lots of works in biomedical tasks~\cite{isensee2021nnu,guan2021multi}. Nevertheless, the target domain is usually inaccessible in the training phase. Domain generalization~\cite{2021Generalizing} aims to improve the generalization to an unseen domain. As a traditional preprocessing method, z-score normalization contributes to the model generalization~\cite{reinhold2019evaluating}. For site-invariant learning, adversarial training~\cite{NEURIPS2018_717d8b3d, zhao2019multi} and meta learning~\cite{li2019feature, liu2021feddg} are both mainstream strategies. Some researchers also enrich the diversity of training data via data augmentation~\cite{yang2022source, zhou2022generalizable}. However, those methods do not consider processing the ATLAS dataset for stroke lesion segmentation, which is composed of MR images from 9 sites.

To overcome the challenges mentioned above, we propose \methodname, a self-adaptive normalization network based on U-net for stroke lesion segmentation. \methodname integrates image-level data harmonization and feature-level site-invariant representation learning to boost the model generalization to unseen sites. Specifically, we present masked adaptive instance normalization (MAIN) prior to feeding MR images to U-net. Although the overall intensity values of MR images from different sites are significantly different, the intensity values of different parts are, by nature, relatively consistent. For instance, the intensity values of white matter are always higher than those of gray matter and lower than those of cerebrospinal fluid in the brain. Enlightened by z-score normalization and dynamic network, we implement MAIN as an adaptive linear transformation. It learns affine parameters from input images and dynamically standardizes input images into a site-unrelated style. In the meanwhile, we leverage the classifier with a gradient reversal layer to guide MAIN and the encoder of U-net to learn site-invariant representation. Note that our SAN is conceptually different from~\cite{yang2022instance} that adaptively uses various normalization techniques. Furthermore, inspired by the ``pseudosymmetry'' structure of the human brain, we also introduce a simple yet effective data augmentation technique called symmetry-inspired data augmentation (SIDA). SIDA is embedded within \methodname to double the number of samples internally (regarding one MR image as one sample for the network), and to halve memory consumption. As a result, stroke lesions from the whole brain can simply be identified within single cerebral hemispheres, making it much easier to locate the lesions.

In summary, the contributions of this paper are listed as follows.
\begin{enumerate}

\item We introduce \methodname, a self-adaptive normalization network based on U-net. To the best of our knowledge, this is the \emph{first} attempt at site generalization for stroke lesion segmentation.
\item Motivated by traditional z-score normalization and dynamic network, masked adaptive instance normalization (MAIN) is designed to dynamically standardize the input MR images into a site-unrelated style. It is coupled with feature-level site-invariant representation learning through a site classifier with a gradient reversal layer.
\item We also introduce a simple yet effective data augmentation technique, termed symmetry-inspired data augmentation (SIDA). As a result, the number of samples is doubled internally (regarding one MR image as one sample for the network), and the memory consumption is halved, making the model locate the lesions much easier.
\item Through extensive experiments on the benchmark Anatomical Tracings of Lesions After Stroke (ATLAS) dataset collected from \emph{nine} different sites, the results show that under the ``leave-one-site-out'' setting, the proposed \methodname outperforms recently published methods. Moreover, the trained models generalize better to unseen external sites than the previous best methods. 

\end{enumerate}

\section{Related Work}\label{sec:rel}

\subsection{Stroke Lesion Segmentation}\label{sec:rel:seg}

Recently, deep learning-based approaches, especially convolutional neural networks (CNNs)~\cite{ 2021Review, qiu2022fgam}, have shown great potential in medical image segmentation.
There are many customized U-nets using attention modules~\cite{ karthik2021delineation, karthik2022contour, hashemi2022delve, karthik2019deep} or multi-scale fusion modules~\cite{karthik2021ischemic}. 
Besides,~\cite{xiong2022weak} combined Bayesian with the traditional U-net architecture, aiming at weak labels instead of full-supervised learning.
We briefly survey works related to stroke lesion segmentation on the benchmark ATLAS dataset~\cite{2018A}.

Given that training the 3D network with a large number of parameters using limited samples can lead to overfitting, some researchers divided MR images into 2D slices and used them as the input to 2D CNNs.
X-Net, which includes a non-local attention module~\cite{2019X}, and MSDF-Net, which involves a multi-scale deep fusion module~\cite{2019MSDF}, were proposed to capture long-range dependencies. Such methods~\cite{2019X, 2019MSDF,yang2019clci} provide more feature relationships for stroke lesion segmentation; nonetheless, 2D CNNs capture few inter-slice dependencies. 
To integrate inter-slice and intra-slice dependencies, 3D CNNs are commonly employed. Some researchers focused on the fusion of 2D and 3D CNNs, such as D-UNet~\cite{2019D}, DFENet~\cite{DBLP:journals/sncs/BasakHR21}, etc.~\cite{2019A, DBLP:conf/miccai/ZhangWLCWT20}. In addition to model architecture, efforts have also been made to improve the training strategy. Based on 3D residual U-net, a zoom-in\&out training strategy~\cite{2020Automatic} was 
designed. It first trained the model on images with small lesions, and then fine-tuned the model on images with large lesions. The input of MI-UNet~\cite{2020MI} was concatenated with brain parcellation as prior knowledge to achieve more accurate stroke segmentation. Nevertheless, the registration step before deep learning requires considerable time cost. Though 3D CNNs generally perform better than 2D CNNs, they typically require more training data to prevent overfitting. The situation becomes even worse when the different site presents different variations in the multi-site scenario.

Therefore, there is an urgent need for stroke lesion segmentation methods that would allow a trained model to adaptively generalize to unseen sites. Though the customized U-nets can provide more precise segmentation, our \methodname focuses on the domain generalization issue. In this case, the conventional U-net is employed as backbone to highlight the effect of each component.

\subsection{Multi-site learning}

In practice, it would be desirable to obtain a trained model that can work well with data from a new hospital or site without further collecting extra data to perform fine-tuning. This is essentially related to multi-site or multi-domain learning in the medical image field, such as domain adaptation~\cite{wang2018deep}. Domain adaptation has been widely used in biomedical tasks~\cite{sun2021multi,isensee2021nnu,guan2021multi}, maximizing the performance on a given target domain. In contrast, domain generalization~\cite{2021Generalizing} aims to maximize the generalization to an unseen domain. This is the problem we are tackling for stroke lesion segmentation.

On the one hand, histogram matching~\cite{A2020Standardization} and z-score normalization~\cite{reinhold2019evaluating} are the most commonly used preprocessing to achieve multi-site learning. Since the training data often consists of multiple sites, histogram matching generally matches the histogram distribution of training data to that of the testing data; however, it requires the latter to be known before training, a problem applicable to domain adaptation but not to domain generalization. Z-score normalization consists of subtracting the mean intensity of the region of interest (covered by the brain mask) from each voxel value and dividing it by the corresponding standard deviation. However, its effectiveness remains limited and is inconsistent across data from different sites.

Adversarial training is one effective strategy to learn domain-invariant representation for multi-site learning. Among several adversarial training, a simple and popular way is to use the gradient reversal layer (GRL)~\cite{ganin2015unsupervised, ganin2016domain}. GRL treats domain invariance as a classification problem and directly maximizes the loss function of the domain classifier by reversing its gradients, which matches the marginal distributions across domains. It has been used widely in various comparative studies~\cite{tsai2018learning, pmlr-v70-long17a, wilson2020survey} and domain adaptation~\cite{NEURIPS2018_717d8b3d, zhao2019multi}. In this paper, the site classifier with a gradient reversal layer is employed to minimize the inter-site diversity.

On the other hand,  various domain generalization methods~\cite{dou2019domain} have been proposed. To reduce the serious domain shift due to nonbiological factors (\eg the acquisition protocols and hardware), Unlearning~\cite{dinsdale2021deep} was a flexible harmonization framework related to adversarial training. Motivated by spectral transfer~\cite{yang2020fda}, FACT~\cite{xu2021fourier} was a Fourier-based data augmentation method that changes the low-frequency amplitude information to improve the model generalization. Then, based on the Fourier-based data augmentation method, RAM-DSIR~\cite{zhou2022generalizable} was proposed as a multi-task paradigm by combining the segmentation model with a self-supervision domain-specific image restoration module for model generalization. Furthermore, for the sake of promoting domain-independent feature cohesion, meta-learning~\cite{li2019feature} could also be combined with episodic learning in continuous frequency space~\cite{liu2021feddg}. More comprehensive surveys about domain generalization can be found in~\cite{2021Generalizing, zhou2022domain}.

In summary, when processing the ATLAS dataset for stroke lesion segmentation, the existing methods typically mix the data from all sites and do not consider the generalization of the trained model to an unseen site. To date, no studies have yet performed leave-one-site-out validation on ATLAS dataset.
In addition, the 9 sites also present challenges for multi-site learning.

\begin{figure*}[t]
\centering
\includegraphics[width=1\linewidth]{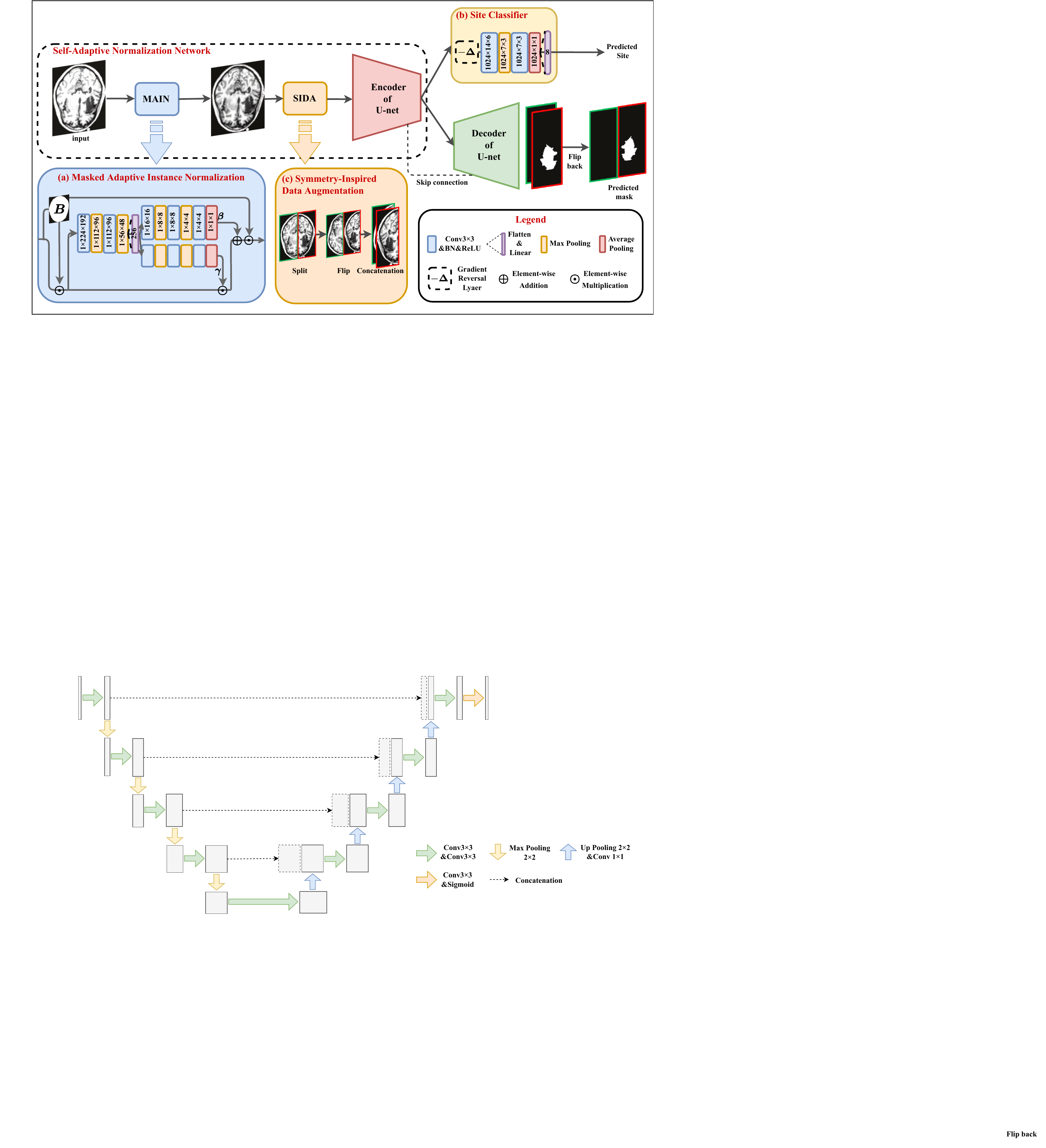}
\caption{Illustration of the proposed \methodname for brain stroke lesion segmentation: (\textbf{a}) Masked Adaptive Instance Normalization (MAIN) standardizes the MR images into a site-unrelated style; (\textbf{b}) Site classifier with gradient reversal layer ($-{\Delta}$) performs site-invariant learning; (\textbf{c}) Symmetry-Inspired Data Augmentation (SIDA) simplifies the training for locating the lesions.}
\label{fig:architecture}
\end{figure*}

\section{Methodology}\label{sec:method}

We present \methodname to improve stroke lesion segmentation model generalization to unseen sites, as illustrated in Fig.~\ref{fig:architecture}. To that end, we use U-net as the backbone and implement MAIN and the site classifier with a gradient reversal layer for site-invariant representation. A simple yet effective data augmentation technique is also introduced. We now describe the key components in detail, followed by the detailed network architecture and the loss function.

\subsection{Preprocessing}

The released ATLAS dataset has been defaced, intensity normalized, and normalized to standard (MNI-152) space (for more details refer to normalization steps in~\cite{2018A}). We found that processing ATLAS dataset by z-score normalization improves the model performance. For a fair comparison, we use z-score normalization to normalize the MR images based on 2D brain slices before the MR images are fed into the networks. 
Specifically, given brain mask $\textbf{B}$ of input MR image $\mat{I}$ of size $w\times h$, we first calculate the mean $\mu^{\text{zs}}$ and the standard deviation $\sigma^{\text{zs}}$ of the intensities within $\mat{B}$. The resulting z-score--normalized image $\mat{I}^{\text{zs}}$ is defined as:
\begin{align}
\mat{I}^{\text{zs}}_{ij} = (\mat{I}_{ij} - \mu^{\text{zs}})/\sigma^{\text{zs}},
\end{align}
where $ij$ represents the position of the pixel in the MR image.

Furthermore, the z-score--normalized preprocessing and our MAIN rely on brain masks. Although the quality of brain masks may affect the results, we emphasize that brain masks can be easily obtained and are of robust quality.

\subsection{Self-Adaptive Normalization}
Our self-adaptive normalization (SAN) is implemented via the proposed image-level masked adaptive instance normalization and learned via the feature-level site-invariant learning, as shown in Fig.~\ref{fig:architecture}. 

\subsubsection{Masked Adaptive Instance Normalization}

Although z-score normalization contributes to the generalization of the model on unseen sites, the effect is limited. The limited generalization of the traditional normalization techniques motivates us to propose MAIN, a dynamic neural network~\cite{han2021dynamic} that can 
adaptively learn how to normalize different input images. MAIN leverages two affine transformation parameters learned dynamically from the masked input MR image to convert z-score--normalized images into a site-unrelated style.

We assume the affine transformation parameters $\gamma$ and $\beta$ are learned from the masked z-score--normalized image $\mat{B}\odot\mat{I}^{\text{zs}}$, where we only care about the brain region. The output of MAIN can be represented as:
\begin{align}
\label{eq}
\mat{I}^{\text{main}}_{ij} = (\gamma \cdot \mat{I}^{\text{zs}}_{ij} + \beta) \cdot \mat{B}_{ij}.
\end{align}

Fig.~\ref{fig:architecture}(a) illustrates the basic structure of the proposed MAIN: a lightweight network with two heads is designed to learn the two affine transformation parameters from the masked brain region. More specifically, using a 256-dimensional vector that supposedly contains the statistical information about the intensity distribution from the brain, the two heads learn how to normalize the MR images into a site-unrelated style by making the tissues of the outputs have similar values. As Section~\ref{detail} shows, our MAIN is a lightweight neural network module and does not impose an excessive computational burden.

Compared with traditional instance normalization, MAIN is more suitable for brain MR images. The mean and the standard deviation are computed in the region of interest formed by the brain mask.
In contrast to AdaIN~\cite{2017Arbitrary} that aligns the mean and the variance of the content features with those of the style features, our affine transformation approach is more generalized. More specifically, AdaIN can be seen as a special case of our method, which requires accessing the images from the target site during training. 

\subsubsection{Site-Invariant Learning}

To minimize the inter-site discrepancy, we use the gradient reversal layer in the site classifier to encourage MAIN and the encoder of U-net. The network architecture of the site classifier with its gradient reversal layer is shown in Fig.~\ref{fig:architecture}(b), and it has previously been known to be effective in domain adaptation fields~\cite{ganin2015unsupervised, NEURIPS2018_717d8b3d}.

The gradient reversal layer acts as an identity transformation during the forward propagation. During the backward propagation, it takes the gradient from the subsequent level and changes its sign (\ie, multiplies it by -1) before passing it to the preceding layer~\cite{ganin2016domain}. In this way, the gradient reversal layer can help the encoder extract site-invariant features and the classifier predicts sites simultaneously. Although adversarial training is widely used in domain adaptation, for our stroke lesion segmentation task with a small amount of data, the gradient reversal layer within the site classifier can be more easily trained than iteration optimization in the GAN framework~\cite{goodfellow2014generative}. We also find that the gradient reversal layer is robust to the site-imbalanced data, which is experimentally validated in Section~\ref{sec:exp:data_imbalanced}.

\subsubsection{Optimization}

\begin{figure}
\centering
\includegraphics[width=0.7\linewidth]{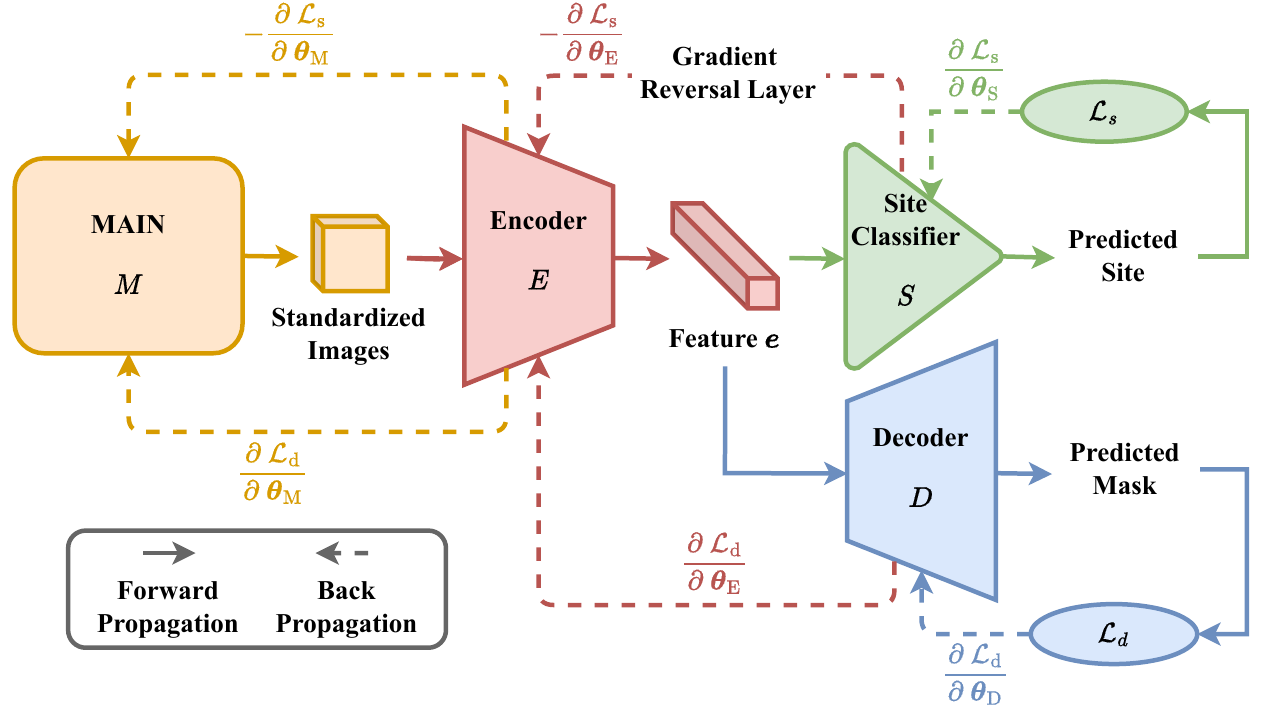}
\caption{Specific implementation of \methodname during training illustrates the overall propagation of the model (dotted arrows denote backpropagation), including MAIN $M$, encoder $E$, decoder $D$, and site classifier $S$. $\mat{\theta}_\text{M}$, $\mat{\theta}_\text{E}$, $\mat{\theta}_\text{D}$, and $\mat{\theta}_\text{S}$ represent the model parameters of these modules, and $\mathcal{L}_\text{d}$ and $\mathcal{L}_\text{s}$ denote the Dice loss and the site loss, respectively.}
\label{fig:grl}
\end{figure}

Fig.~\ref{fig:grl} shows the overall forward and backward propagation during training. The overall training process includes MAIN (orange) and the encoder (pink) and decoder (blue) of U-net, which together form a standard feed-forward architecture. Considering site-related representations influence the segmentation results, domain (site) generalization is achieved by adding a site classifier (green) to MAIN and the encoder via a gradient reversal layer~\cite{ganin2015unsupervised} that multiplies the gradient by a certain negative constant ($-1$ in this paper) during backward propagation-based training. Moreover, the training proceeds in a standard way and minimizes the segmentation Dice loss and the classification site loss.

In this case, the model parameters are optimized with backpropagation:
\begin{align}
&\mat{\theta}_\text{M} \gets \mat{\theta}_\text{M} - \mu \left(\frac{\partial \mathcal{L}_\text{d}}{\partial \mat{\theta}_\text{M}} - \frac{\partial \mathcal{L}_\text{s}}{\partial \mat{\theta}_\text{M}}\right),&&\ 
\mat{\theta}_\text{E} \gets \mat{\theta}_\text{E} - \mu \left(\frac{\partial \mathcal{L}_\text{d}}{\partial \mat{\theta}_\text{E}} - \frac{\partial \mathcal{L}_\text{s}}{\partial \mat{\theta}_\text{E}}\right),\\
&\mat{\theta}_\text{D} \gets \mat{\theta}_\text{D} - \mu \frac{\partial \mathcal{L}_\text{d}}{\partial \mat{\theta}_\text{D}}, &&\ 
\mat{\theta}_\text{S} \gets \mat{\theta}_\text{S} - \mu \frac{\partial \mathcal{L}_\text{s}}{\partial \mat{\theta}_\text{S}},
\end{align}
where $\mu$ is the learning rate; $\mat{\theta}_\text{M}$, $\mat{\theta}_\text{E}$, $\mat{\theta}_\text{D}$, and $\mat{\theta}_\text{S}$ represent the model parameters of MAIN, encoder, decoder, and site classifier, respectively;  $\mathcal{L}_\text{d}$ and $\mathcal{L}_\text{s}$ denote Dice loss and site loss, respectively.

The gradient reversal layer ensures that the feature distributions among the multiple sites are similar (as indistinguishable as possible for the site classifier). This layer also helps MAIN to compute dynamic affine parameters adaptively for every input image. In addition, the proposed MAIN and site-invariant learning help the encoder of U-net extract site-invariant features. Therefore, the skip connections reuse the features from the encoder of U-net, which are supposed to contain little site information.

\subsection{Symmetry-Inspired Data Augmentation}

Although the training of \methodname ensures that the model learns a site-unrelated style of MR images and site-invariant representations, the small number of the training samples in this study limit the generalization, particularly in cases when strong variations in the shape, size, and location of the stroke lesions are present.

Inspired by the ``pseudosymmetry'' of the human brain, it is possible to make locating stroke lesions much easier for the model. There have been some studies that incorporate the symmetry analytical shape as prior knowledge with
machine learning techniques to detect brain abnormalities~\cite{narkhede2014brain, kermi2018fully, liu2019using}.

However, this strategy would increase the complexity of networks and still require locating lesions from the whole image. In addition, some lesion-symptom mapping studies often flip the lesions to one hemisphere to increase the sample number for statistical evaluations~\cite{griffis2016voxel}. It would change the location of lesions, which might be inconsistent with clinical practice.

Unlike the existing methods that use symmetry priors for the whole brain image or flip the lesions to the other hemisphere, we propose a simple yet effective data augmentation method, termed symmetry-inspired data augmentation (SIDA), which is shown in Fig.~\ref{fig:architecture}(c). It can be embedded within the \methodname and that enables \methodname to quickly locate the stroke lesion on brain hemispheres through repetition. We also note that SIDA relies on symmetric body parts (\eg human brain) and spatial normalization preprocessing. In the released dataset, the MR images have been spatially normalized through the MINC-toolkit~\cite{sled1998nonparametric}. Even if the MR images are based on non-symmetric body parts or unsuccessful spatial normalization, the proposed data augmentation could be regarded as a general splitting-based data augmentation. Besides, there is no patient whose lesion spans the hemispheres in ATLAS dataset~\cite{2018A}. If this situation occurs, the left and right halves could be concatenated together to form the complete lesion.

\begin{table}[!t]
\renewcommand{\arraystretch}{1.0}
\caption{Details on the structure of U-net.}
\label{tab:UNet}
\centering
\begin{tabular}{rcl}

\shline
\bfseries \ &\bfseries Feature size &\bfseries Parameters \\

\midrule
Input & \hspace{3.75mm}1 $\times$ 224 $\times$ 96 & \\
\midrule
Conv 1 & \hspace{2.5mm}64 $\times$ 224 $\times$ 96 & [3 $\times$ 3, 64 conv] $\times$ 2$^a$ \\
Pooling & \hspace{2.5mm}64 $\times$ 112 $\times$ 48 & [2 $\times$ 2, max pooling]$^b$ \\
\midrule
Conv 2 & \hspace{1.25mm}128 $\times$ 112 $\times$ 48 & [3 $\times$ 3, 128 conv] $\times$ 2 \\
Pooling & \hspace{1.25mm}128 $\times$ \hspace{1.25mm}56 $\times$ 24 & [2 $\times$ 2, max pooling] \\
\midrule
Conv 3 & \hspace{1.25mm}256 $\times$ \hspace{1.25mm}56 $\times$ 24 & [3 $\times$ 3, 256 conv] $\times$ 2 \\
Pooling & \hspace{1.25mm}256 $\times$ \hspace{1.25mm}28 $\times$ 12 & [2 $\times$ 2, max pooling] \\
\midrule
Conv 4 & \hspace{1.25mm}512 $\times$ \hspace{1.25mm}28 $\times$ 12 & [3 $\times$ 3, 512 conv] $\times$ 2 \\
Pooling & \hspace{1.25mm}512 $\times$ \hspace{1.25mm}14 $\times$ \hspace{1.25mm}6 & [2 $\times$ 2, max pooling] \\
\midrule
Conv 5 & 1024 $\times$ \hspace{1.25mm}14 $\times$ \hspace{1.25mm}6 & [3 $\times$ 3, 1024 conv] $\times$ 2 \\
\midrule
Upsampling & 1024 $\times$ \hspace{1.25mm}28 $\times$ 12 & [2 $\times$ 2, upsampling]-[Conv 4]$^c$ \\
Conv 6 & \hspace{1.25mm}512 $\times$ \hspace{1.25mm}28 $\times$ 12 & [3 $\times$ 3, 512 conv] $\times$ 2 \\
\midrule
Upsampling & \hspace{1.25mm}512 $\times$ \hspace{1.25mm}56 $\times$ 24 & [2 $\times$ 2, upsampling]-[Conv 3] \\
Conv 7 & \hspace{1.25mm}256 $\times$ \hspace{1.25mm}56 $\times$ 24 & [3 $\times$ 3, 256 conv] $\times$ 2 \\
\midrule
Upsampling & \hspace{1.25mm}256 $\times$ 112 $\times$ 48 & [2 $\times$ 2, upsampling]-[Conv 2] \\
Conv 8 & \hspace{1.25mm}128 $\times$ 112 $\times$ 48 & [3 $\times$ 3, 128 conv] $\times$ 2 \\
\midrule
Upsampling & \hspace{1.25mm}128 $\times$ 224 $\times$ 96 & [2 $\times$ 2, upsampling]-[Conv 1] \\
Conv 9 & \hspace{2.5mm}64 $\times$ 224 $\times$ 96 & [3 $\times$ 3, 64 conv] $\times$ 2 \\
\midrule
Output & \hspace{3.75mm}1 $\times$ 224 $\times$ 96 & [1 $\times$ 1, 1 conv]+Sigmoid \\

\shline
\end{tabular}

\begin{flushleft}
\footnotesize{$^a$[3 $\times$ 3, 64 conv] corresponds to a convolution with a kernel size of 3 $\times$ 3 and 64 feature maps.} \\

\footnotesize{$^b$[2 $\times$ 2, max pooling] denotes max pooling with a kernel size of 2 $\times$ 2.} \\

\footnotesize{$^c$[2 $\times$ 2, upsampling] indicates that the height and width of the feature map are twice as large as the original via upsampling, following a convolution with a kernel size of 1 $\times$ 1 and a half as many feature maps; [ ]-[ ] denotes the concatenation of two feature maps.}
\end{flushleft}
\end{table}

\subsection{Network Architecture}
\label{detail}

\subsubsection{MAIN}

As shown in Fig.~\ref{fig:architecture}(a), the branch structures for computing $\gamma$ and $\beta$ are the same. Every convolutional block consists of a 3$\times$3 convolution, batch normalization, and ReLU activation, with one filter. Each max-pooling layer divides the resolution of the feature map by 2$\times$2. The fully connected layer has 2,688 input features and 256 output features, which summarizes the statistics of the input MR images. Then, the outputs from the fully-connected layer are reshaped to size 16$\times$16. The average pooling layer compresses the resolution to 1$\times$1, allowing us to obtain the adaptive affine parameters $\gamma$ and $\beta$ for each input image.

\subsubsection{U-net}

In clinical practice, the number of training samples is often limited. Since training the 3D network with large parameters using limited samples might easily lead to overfitting, we choose the 2D U-net.
As shown in Fig.~\ref{fig:architecture}, the backbone of the model is U-net~\cite{2015U}. The encoder is used for site-invariant representation learning and the decoder is used for segmentation output. In our experiment, one convolutional block consists of a 3$\times$3 convolution, batch normalization, and ReLU activation. Specifically, the structural details of U-net are shown in Table~\ref{tab:UNet}. 

\subsubsection{Site Classifier}

Fig.~\ref{fig:architecture}(b) shows the lightweight structure of the site classifier: two convolutional blocks, one max-pooling layer, one average-pooling layer, one fully connected layer with 1,024 input features, and 8 output features. One convolutional block consists of a 3$\times$3 convolution, batch normalization, and ReLU activation, and both convolutional blocks contain 1,024 filters. The output of the site classifier has 8 classes, corresponding to the 8 sites in the training set.

\subsection{Loss Functions}

We briefly introduce the loss functions used in our experiments. Dice loss~\cite{milletari2016v} to train the segmentation capacity can be formally defined for one sample as follows:
\begin{align}
\mathcal{L}_{\text{d}}=1-\frac{2\sum_{i=1}^{w}\sum_{j=1}^{h} p_{ij} \hat{p}_{ij} + \epsilon}{\sum_{i=1}^{w}\sum_{j=1}^{h} p_{ij}^{2} + \sum_{i=1}^{w}\sum_{j=1}^{h} \hat{p}_{ij}^{2} + \epsilon},
\end{align}
where $p_{ij}$ represents ground truth; $\hat{p}_{ij}$ represents model prediction probability; $\epsilon=10^{-5}$ is a smoothing term.

Then, cross-entropy loss~\cite{murphy2012machine} is used to train the site classifier with a gradient reversal layer in our experiment. Site loss can be formally defined for one sample as follows:
\begin{align}
\mathcal{L}_{\text{s}}=- \sum_{k=1}^{K} y_{k} \log \hat{y}_k ,
\end{align}
where $K$ denotes the number of sites; $y_{k}$ represents whether the sample comes from the $k$-th site; and $\hat{y}_{k}$ represents the site classifier prediction probability value for the $i$-th site.

Our goal is to train a segmentation model that not only performs well but also generalizes well to unseen sites. Therefore, the total loss function for optimizing \methodname is defined for one sample as follows:
\begin{align}
\mathcal{L}=\mathcal{L}_{\text{d}}+ \mathcal{L}_{\text{s}}.
\end{align}

\section{Experiment}\label{sec:exp}

We empirically evaluate \methodname on the benchmark stroke lesion dataset ATLAS, composed of T1-weighted MR images from 9 different sites. We start by describing the benchmark ATLAS in Section~\ref{sec:exp:dataset} and the implementation details in Section~\ref{sec:exp:details}, followed by details of an ablation study in Section~\ref{sec:exp:ablation} and a comparison with recently published methods under the ``leave-one-site-out''~\cite{esteban2017mriqc} setting in Section~\ref{sec:exp:comparison}. Finally, further evaluation and analysis on 
the generalization to unseen external sites, data imbalance across sites, the performance of individual sites, and robustness verification on the midline shift are presented in Section~\ref{sec:exp:further_results}. 

\newcommand{\tabincell}[2]{
\begin{tabular}{@{}#1@{}}#2\end{tabular}
}

\begin{table*}[!t]
\renewcommand{\arraystretch}{1.0}
\caption{The nine source sites of the T1-weighted MR images in our experiment.}
\label{tab:site}
\centering
\begin{tabular}{clllr}

\shline
&\textbf{Site} &\textbf{Location} & \textbf{Scanner} & \# \textbf{Patients} \\

\midrule
1 & \tabincell{l}{Medical University General Hospital} & \tabincell{l}{Tianjin, China} & \tabincell{l}{GE 750 Discovery} & 55\\
2 & University of Tübingen & \tabincell{l}{Tübingen, Germany} & \tabincell{l}{GE Signa Excite} & 34\\
3 & \tabincell{l}{Sunnaas Rehabilitation Hospital} & \tabincell{l}{Nesodden, Norway} & \tabincell{l}{Siemens Trio} & 27\\
4 & \tabincell{l}{NORMENT and KG Jebsen Centre \\for Psychosis Research} & \tabincell{l}{Oslo, Norway} & \tabincell{l}{Siemens Trio} & 12\\
5 & \tabincell{l}{Department of Psychology} & \tabincell{l}{Oslo, Norway} & \tabincell{l}{Phillips Achieva} & 27\\
6 & Child Mind Institute & \tabincell{l}{New York, USA} & \tabincell{l}{Siemens Trio} & 14\\
7 & \tabincell{l}{Nathan S. Kline Institute \\for Psychiatric Research} & \tabincell{l}{Orangeburg, USA} & \tabincell{l}{Siemens Trio} & 11\\
8 & \tabincell{l}{University of Texas Medical Branch} & \tabincell{l}{Galveston, USA} & \tabincell{l}{GE 750 Discovery} & 35\\
9 & University of Michigan & \tabincell{l}{Ann Arbor, USA} & \tabincell{l}{Siemens Trio} & 14\\
\shline
\end{tabular}
\end{table*}

\subsection{ATLAS Dataset}\label{sec:exp:dataset}

ATLAS dataset has two versions: ATLAS v1.2~\cite{2018A} and ATLAS v2.0~\cite{Liew2021.12.09.21267554}.
The existing methods were mainly evaluated on ATLAS v1.2, which has been well-studied. If we compare the proposed \methodname with other methods using ATLAS v2.0~\cite{Liew2021.12.09.21267554}, the parameter settings of other methods may not be optimal. For a fair comparison, our ablation study and comparison experiments are mainly based on ATLAS v1.2. In addition, two sites from ATLAS v2.0 were selected for external evaluation.

\begin{table}[h]
\renewcommand{\arraystretch}{1.0}
\caption{An ablation study on ATLAS v1.2 for various evaluation metrics (the standard deviation behind $\pm$ is across subjects). SL: Site-invariant learning. Since all p-values in Wilcoxon tests are less than $5$\%, we do not show them in the table.}
\label{tab:ablation}
\centering
\begin{tabular}{cccccc}

\shline

\bfseries SIDA &\bfseries SL &\bfseries MAIN &\bfseries Dice &\bfseries Recall &\bfseries F1-score\\

\midrule

\multicolumn{6}{c}{\textbf{Site 1 / GE 750 Discovery / 55 Subjects}}\\ \hline

\ & \ & \ & 0.5295{\scriptsize$\pm$0.2407} &  0.4416$\pm${\scriptsize0.2015} & 0.4565{\scriptsize$\pm$0.2406} \\
\checkmark & \ & \ & 0.5436{\scriptsize$\pm$0.2134} &  0.4560{\scriptsize$\pm$0.2141} & 0.4827{\scriptsize$\pm$0.2134}\\
\ & \checkmark & \ & 0.5626{\scriptsize$\pm$0.2018}  & 0.4907{\scriptsize$\pm$0.2242} & 0.5335{\scriptsize$\pm$0.2018}\\
\checkmark & \checkmark & \ & 0.5866{\scriptsize$\pm$0.2274}  & 0.6048{\scriptsize$\pm$0.2315} & 0.6077{\scriptsize$\pm$0.2072}\\
\ & \checkmark & \checkmark & 0.5930{\scriptsize$\pm$0.2292} &  0.5729{\scriptsize$\pm$0.2429} & 0.6154{\scriptsize$\pm$0.2292}\\
\checkmark & \checkmark & \checkmark & \textbf{0.6018}{\scriptsize$\pm$0.2344}  & \textbf{0.6570}{\scriptsize$\pm$0.2354} & \textbf{0.6246}{\scriptsize$\pm$0.2276}\\

\midrule
\multicolumn{6}{c}{\textbf{Site 4 / Siemens Trio / 12 Subjects}}\\ \hline

\ & \ & \ & 0.2928{\scriptsize$\pm$0.2846} & 0.3658{\scriptsize$\pm$0.3186} & 0.4437{\scriptsize$\pm$0.3182}\\
\checkmark & \ & \ & 0.3130{\scriptsize$\pm$0.2906} & 0.3576{\scriptsize$\pm$0.3186} & 0.4529{\scriptsize$\pm$0.3259}\\
\ & \checkmark & \ & 0.3011{\scriptsize$\pm$0.2796}  & 0.3714{\scriptsize$\pm$0.3063} & 0.4577{\scriptsize$\pm$0.3017} \\
\checkmark & \checkmark & \ & 0.3322{\scriptsize$\pm$0.2916}  & 0.3783{\scriptsize$\pm$0.3259} & 0.4691{\scriptsize$\pm$0.3172}\\
\ & \checkmark & \checkmark & 0.3529{\scriptsize$\pm$0.2865}  & 0.4128{\scriptsize$\pm$0.3182} & 0.4706{\scriptsize$\pm$0.3276} \\
\checkmark & \checkmark & \checkmark & \textbf{0.3625}{\scriptsize$\pm$0.2893}   & \textbf{0.4439}{\scriptsize$\pm$0.3042} & \textbf{0.4883}{\scriptsize$\pm$0.3126}\\
\midrule

\multicolumn{6}{c}{\textbf{Site 5 / Phillips Achieva / 27 Subjects}}\\ \hline

\ & \ & \ & 0.4396{\scriptsize$\pm$0.2263} & 0.4044{\scriptsize$\pm$0.2631} & 0.4565{\scriptsize$\pm$0.2263}\\
\checkmark & \ & \ & 0.4659{\scriptsize$\pm$0.2605} & 0.4365{\scriptsize$\pm$0.2870} & 0.4831{\scriptsize$\pm$0.2605}\\
\ & \checkmark & \ & 0.4766{\scriptsize$\pm$0.2199} & 0.5138{\scriptsize$\pm$0.2615} & 0.4966{\scriptsize$\pm$0.2201} \\
\checkmark & \checkmark & \ & 0.4946{\scriptsize$\pm$0.2472}& 0.5268{\scriptsize$\pm$0.2814} & 0.5335{\scriptsize$\pm$0.2472}\\
\ & \checkmark & \checkmark & 0.5123{\scriptsize$\pm$0.2108} & 0.5069{\scriptsize$\pm$0.2530} & 0.5123{\scriptsize$\pm$0.2108}\\
\checkmark & \checkmark & \checkmark & \textbf{0.5322}{\scriptsize$\pm$0.2474} & \textbf{0.5294}{\scriptsize$\pm$0.2794} & \textbf{0.6040}{\scriptsize$\pm$0.2474}\\

\shline
\end{tabular}
\end{table}

ATLAS v1.2 is an open-source dataset, containing T1-weighted MR images of 229 defaced patients with normalized intensity localized in the standard space (MNI-152 template). Actually, the size of the original dataset is 309. Due to technical difficulties and differences in scanner image quality, the original authors of ATLAS v1.2 excluded the images of lower resolution or failed registration~\cite{2018A}. Following~\cite{2019X, 2019MSDF,yang2019clci, 2019A, 2019D}, we use the MR images of 229 patients. Each patient has $189$ T1-weighted normalized image slices of size $233 \times 197$.  Table~\ref{tab:site} presents the site information, location, scanner, and the number of patients for 9 sites. There is no test-retest scan within the site or between sites, except Site 8 which includes 9 test-retest scans (regarded as 18 patients). It has no effect on the ``leave-one-site-out'' validation, as the test-retest scans only exist on one site. The large intersite discrepancy is described as follows. First, this dataset covers four countries and eight cities, which leads to different vascular territories due to population variation. Second, the ATLAS dataset is acquired by different 3T MR scanners, including GE 750 Discovery, GE Signa Excite, GE Signa HD-X, Phillips Achieva, and Siemens Trio. Third and most importantly, diverse imaging protocols are used even on the same scanner. These factors lead to substantial challenges in generalizing across sites.

The lesion masks were manually segmented by two specialists collaboratively, and quality control was performed on each lesion mask by a second tracer. Consequently, these manual lesion masks serve as ground truth. The lesions range in size from $10$ to $2.838\times10^5$ mm$^3$. As for location, most lesions are roughly equally distributed between the left and right hemispheres ($48$\% left hemisphere vs $44$\% right hemisphere); only $8$\% are found in other locations, such as the brainstem and the cerebellum, which supports the rationale of our proposed data augmentation method. Overall, slightly more than half of the subjects have only one lesion~($58$\%), while the remaining have multiple lesions~($42$\%).

\subsection{Implementation Details}\label{sec:exp:details}

The proposed \methodname is implemented in PyTorch\footnote{The source code is available at~\url{https://github.com/wyyu0831/SAN}.} and trained on a single NVIDIA V100 Tensor Core GPU. The sampling procedure is random selection using all slices. In our model, we use stochastic gradient descent to optimize the trainable parameters, with an initial learning of 0.001 and a 4\% weight decay after each epoch for a total of 50 epochs. The mini-batch size is set to 16 for training. We initially validate the performance of the model on Site 5 and use the same configuration for all sites. Following~\cite{2019X,DBLP:conf/miccai/ZhangWLCWT20}, all slices are cropped into a size of 224$\times$192.

\subsection{Ablation Study}\label{sec:exp:ablation}

To prove the contribution of each component, we take U-net and the z-score--normalized MR images as the baseline. We select three sites with different types of scanners as the testing sets (Sites 1, 4, and 5) for the ablation study. The segmentation results for various component combinations are presented in Table~\ref{tab:ablation}. In the meanwhile, we adopt Wilcoxon test~\cite{conover1999practical} to examine if each combination of the components has the same performance as the baseline. It shows each component consistently contributes to an increasing segmentation performance across different sites.

We note that z-score normalization also contributes to site generalization. Taking Site 5 as an example, if we feed the U-net with unnormalized MR images, the segmentation performance drops substantially (Dice: $0.3794$, Recall: $0.3408$, F1-score: $0.3746$). In other words, z-score normalization increases the Dice coefficient by 0.06. Whereas, the generalization effect of z-score normalization remains limited. The following analysis demonstrates that our SAN-Net achieves a better generalization on unseen sites.

\begin{figure}
\centering
\includegraphics[width=0.45\linewidth]{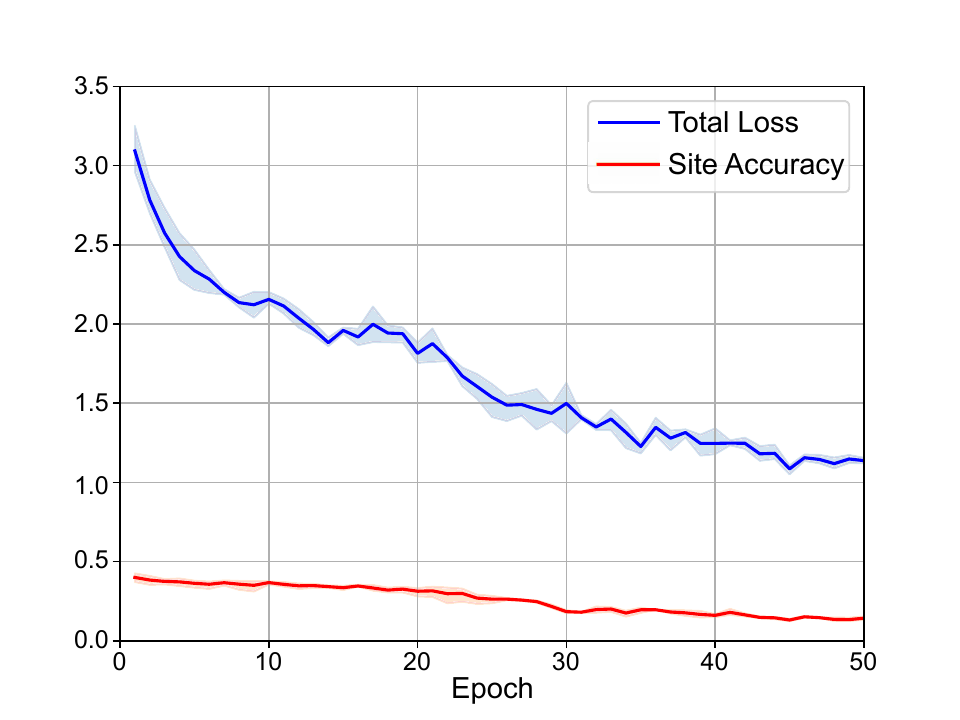}
\caption{Training curves of the total loss for the whole model and the site accuracy for the classifier during training.}
\label{fig:training}
\end{figure}

\begin{figure}[h]
\centering
\includegraphics[width=1\linewidth]{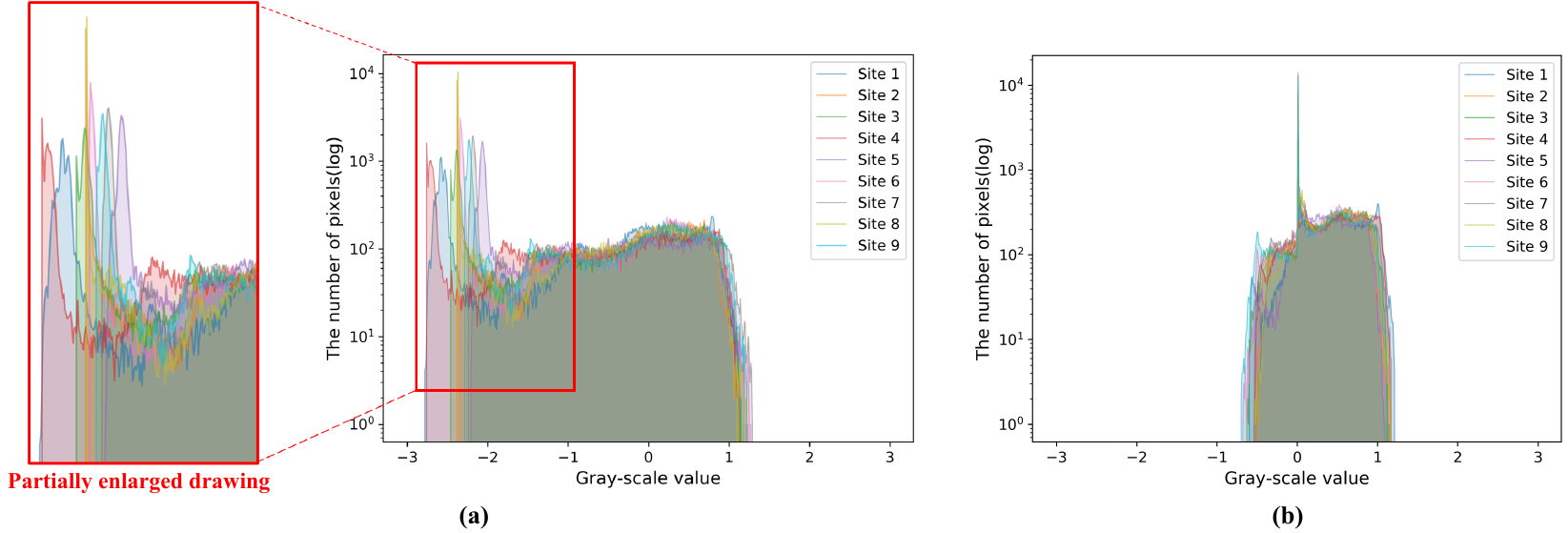}
\caption{The distribution of grayscale values of the same slice in nine different sites processed by (\textbf{a}) traditional z-score normalization, and (\textbf{b}) masked adaptive instance normalization.}
\label{fig:histogram}
\end{figure}

\begin{figure}
\centering
\includegraphics[width=0.45\linewidth]{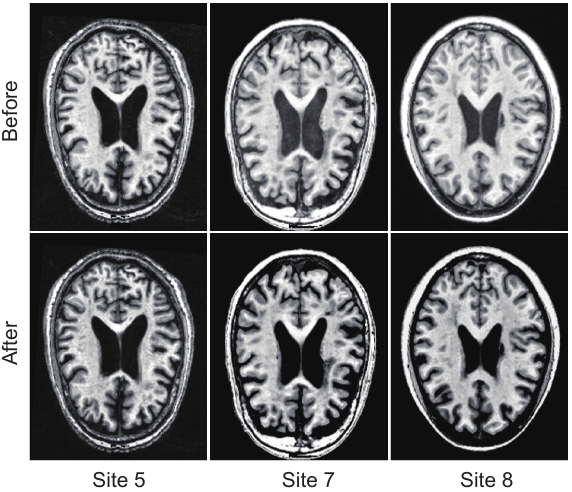}
\caption{Examples of MR slices before and after MAIN. From left to right, the three columns correspond to Sites 5, 7, and 8.}
\label{fig:instance}
\end{figure} 

\subsubsection{Effect of Site-Invariant Learning}\label{sec:exp:invariant}

Taking Site 5 as an example, Table~\ref{tab:ablation} shows that the site classifier with a gradient reversal layer increases the Dice coefficient from 0.4396 to 0.4766 for the baseline U-net, and from 0.4695 to 0.4946 for the baseline U-net with data augmentation. Thus, the gradient reversal layer contributes to learning site-invariant representations for improving generalization to unseen sites. We repeat the same training process five times and display the averaged training curves with a standard deviation as error bound in Fig.~\ref{fig:training}, which shows that the model converges well. More importantly, the accuracy of the site classifier reaches approximately 0.14---close to an ideal random guess ($\tfrac{1}{8}=0.125$). The site classifier cannot identify which site the sample comes from, contributing to site-invariant learning.

\begin{table}[h]
\centering
\caption{Comparison of the metrics of our model without and with SIDA.}
\label{tab:da}
\begin{tabular}{lccc}

\shline
& \textbf{Memory} [MB] & \textbf{MACC} [G] & \textbf{FLOPs} [G] \\

\midrule
\methodname w/o SIDA & 263.52 & 67.17 & 33.61 \\

\methodname w/\hspace{1.5mm} SIDA & 132.11 & 33.59 & 33.61 \\

\shline
\end{tabular}
\end{table}

\subsubsection{Effect of MAIN}

Since MAIN should be combined with site-invariant learning to learn a site-unrelated style, we do not list the results of MAIN only. On the grounds of Table~\ref{tab:ablation}, MAIN also consistently improves all metrics. 
Furthermore, we take the 94th slice among 189 slices for every patient from each site and plot the distributions of their grayscale values before and after the application of MAIN. According to Fig.~\ref{fig:histogram}(a), only traditional z-score normalization is incapable of managing the misalignment among different sites for low grayscale value regions, which is highlighted in the external partially enlarged drawing.
In contrast, Fig.~\ref{fig:histogram}(b) shows that with MAIN, the resulting distributions from different sites are quite similar to each other, potentially eliminating the discrepancies in image appearance. 

Because of the brain mask $\mat{B}$ in Eq.~\eqref{eq}, there is a peak at a grayscale value of zero after applying MAIN, which increases the normality of the distribution. As a result, MAIN leads to better generalization to unseen sites by dynamically standardizing the input MR images. It is worth noting that the z-score normalization is performed slice-wise within the range of the brain mask. Nevertheless, the histogram distribution is based on both the brain mask and the background, leading to a mean value of nonzero in Fig.~\ref{fig:histogram}. In addition, we randomly select three patients from three different sites and show the three corresponding 94th slices before and after MAIN in Fig.~\ref{fig:instance}. Intuitively, MAIN makes the three MR images look more similar.

\subsubsection{Effect of SIDA}

The proposed data augmentation method SIDA consistently improves all metrics by large margins, as shown in Table~\ref{tab:ablation}. Taking Site 5 for example, the Dice coefficient is increased by approximately 0.03, demonstrating the effectiveness of SIDA. Table~\ref{tab:da} also shows that SIDA halves the memory consumption of the model and doubles the computational speed in terms of multiply-accumulate operations (MACC). The \methodname with SIDA has the same floating-point operations (FLOPs) as the one without it, as \methodname requires performing segmentation on two cerebral hemispheres. To summarize, the proposed SIDA improves the memory usage and computational efficiency of the model and makes it much easier for locating stroke lesions.

\subsection{Comparison to Prior Work}\label{sec:exp:comparison}

Unlike the existing works that partition the ATLAS dataset via patient indexes, we directly follow the ``leave-one-site-out'' principle~\cite{esteban2017mriqc}, \ie, taking images obtained at one site as the testing set and those at the other eight sites as the training set. By repeating this process 9 times, we evaluate our \methodname in terms of quantitative metrics and visual analysis. The prior work we compared includes U-net~\cite{2015U}, 
U-Net3+~\cite{DBLP:conf/icassp/HuangLTHZIHCW20}, nnU-Net~\cite{isensee2021nnu}, X-Net~\cite{2019X},  CLCI-Net~\cite{yang2019clci}, DFENet~\cite{DBLP:journals/sncs/BasakHR21}, FACT~\cite{xu2021fourier}, Unlearning~\cite{dinsdale2021deep}, and RAM-DSIR~\cite{zhou2022generalizable}. 

Following~\cite{tustison2013instrumentation}, we reproduce baseline methods as follows. First, the structure of U-net and ResUNet baseline is based on the structure of U-net we used in Table~\ref{tab:UNet}.  U-Net3+, nnU-Net, X-Net, CLCI-Net, FACT, Unlearning, and RAM-DSIR are implemented via the original authors' GitHub codes. Moreover, DFENet is implemented on the grounds of the original authors' description. Second, all baseline methods follow the original authors' default settings. Third, the preprocessing steps for those models are the same. On the basis of the preprocessing steps already done in the released dataset, the other methods take z-score--normalized MR images as input.  

\begin{table*}[!t]
\renewcommand{\arraystretch}{1.0}
\caption{Comparison with other methods in terms of various evaluation metrics (the standard deviation behind ± is across nine experiments). \#Par.: the number of model parameters [M]; Mem.: total GPU memory of the model [MB]; MACC: multiply-accumulate operations [G]; and FLOPs: floating-point operations [G].}
\label{tab:performance_comparison}
\centering
\resizebox{1.0\linewidth}{!}{
\begin{tabular}{rcccrrrr}
\shline
\textbf{Method} &\textbf{Dice} &\textbf{Recall} &\textbf{F1-score} & \#\textbf{Par.}  & \textbf{Mem.}  & \textbf{MACC}  & \textbf{FLOPs} \\

\midrule
U-net~\cite{2015U} & 0.4712{\scriptsize$\pm$0.1952} & 0.4315{\scriptsize$\pm$0.1931} & 0.4864{\scriptsize$\pm$0.2161} & 28.94 & 260.20 & 63.21 & 31.63\\

U-Net3+~\cite{DBLP:conf/icassp/HuangLTHZIHCW20} & 0.5210{\scriptsize$\pm$0.2077} & 0.4851{\scriptsize$\pm$0.1849} & 0.4972{\scriptsize$\pm$0.1930} & 26.97 & 961.57 & 259.57 & 129.87\\

nnU-Net~\cite{isensee2021nnu} & 0.5047{\scriptsize$\pm$0.2002} & 0.4916{\scriptsize$\pm$0.1990} & 0.5268{\scriptsize$\pm$0.2026} & 18.67 & 155.01 & \textbf{21.22} & 10.18\\

X-Net~\cite{2019X} & 0.5083{\scriptsize$\pm$0.1926} & 0.4954{\scriptsize$\pm$0.1844} & 0.5179{\scriptsize$\pm$0.1896} & 15.05 & 915.67 & 40.49 & 20.33\\

CLCI-Net~\cite{yang2019clci} & 0.5174{\scriptsize$\pm$0.1928} & 0.5139{\scriptsize$\pm$0.1975} & 0.5128{\scriptsize$\pm$0.1838} & 36.81 & 1235.35 & 35.71 & \textbf{8.0}\\

DFENet~\cite{DBLP:journals/sncs/BasakHR21} & 0.5302{\scriptsize$\pm$0.2029} & 0.5457{\scriptsize$\pm$0.1874} & 0.5266{\scriptsize$\pm$0.1947} & 16.72 & 1083.52 & 58.63 & 27.49\\

FACT~\cite{xu2021fourier} & 0.5488{\scriptsize$\pm$0.1915} & 0.5583{\scriptsize$\pm$0.1744} & 0.5563{\scriptsize$\pm$0.1934} & 28.94 & 260.20 & 63.21 & 31.63\\

Unlearning~\cite{dinsdale2021deep} & 0.5415{\scriptsize$\pm$0.1881} & 0.5632{\scriptsize$\pm$0.1721} & 0.5365{\scriptsize$\pm$0.1881} & 27.90 & 205.73 & 50.50 & 23.86\\

RAM-DSIR~\cite{zhou2022generalizable} & 0.5562{\scriptsize$\pm$0.1906} & 0.5674{\scriptsize$\pm$0.1835} & 0.5482{\scriptsize$\pm$0.1963} & \textbf{10.59} & 273.24 & 21.24 & 10.65\\

\methodname (ours) & \textbf{0.5711}{\scriptsize$\pm$0.1957} & \textbf{0.5977}{\scriptsize$\pm$0.1587} & \textbf{0.5623}{\scriptsize$\pm$0.1926} & 29.64 & \textbf{130.79} & 31.60 & 33.63 \\

\shline

\end{tabular}}
\end{table*}

\subsubsection{Quantitative Comparison}

Table~\ref{tab:performance_comparison} reports the mean and the standard deviation of the performance on the 9 testing sets, which demonstrates that our method has better segmentation performance in all metrics. In the testing phase, the proposed \methodname has competitive model memory, FLOPs, and MACC as other methods. In other words, our framework ensures generalized performance without the extra GPU and other computational costs compared to its baseline U-net. For instance, RAM-DSIR~\cite{zhou2022generalizable} is a recently proposed image-level and feature-level generalization learning method. Domain-specific batch normalization (DSBN) layers, a component of RAM-DSIR, are suitable for the situation where one patient is only corresponding to one slice. For the ATLAS dataset, one patient involves multiple brain slices of different positions, which is unfriendly to DSBN layers without the corresponding brain masks.

We note that the preprocessing time should be considered in clinical applications, yet the preprocessing has been done in the released dataset. The inference time in Table~\ref{tab:performance_comparison} only refers to the network processing, which is fair for all compared methods. We also adopt Wilcoxon test~\cite{conover1999practical} to confirm the statistical significance.
Therefore, Table~\ref{tab:performance_comparison} shows the superiority of our framework over recently published methods in terms of quantitative metrics.

\begin{figure*}[!t]
\centering
\includegraphics[width=1\linewidth]{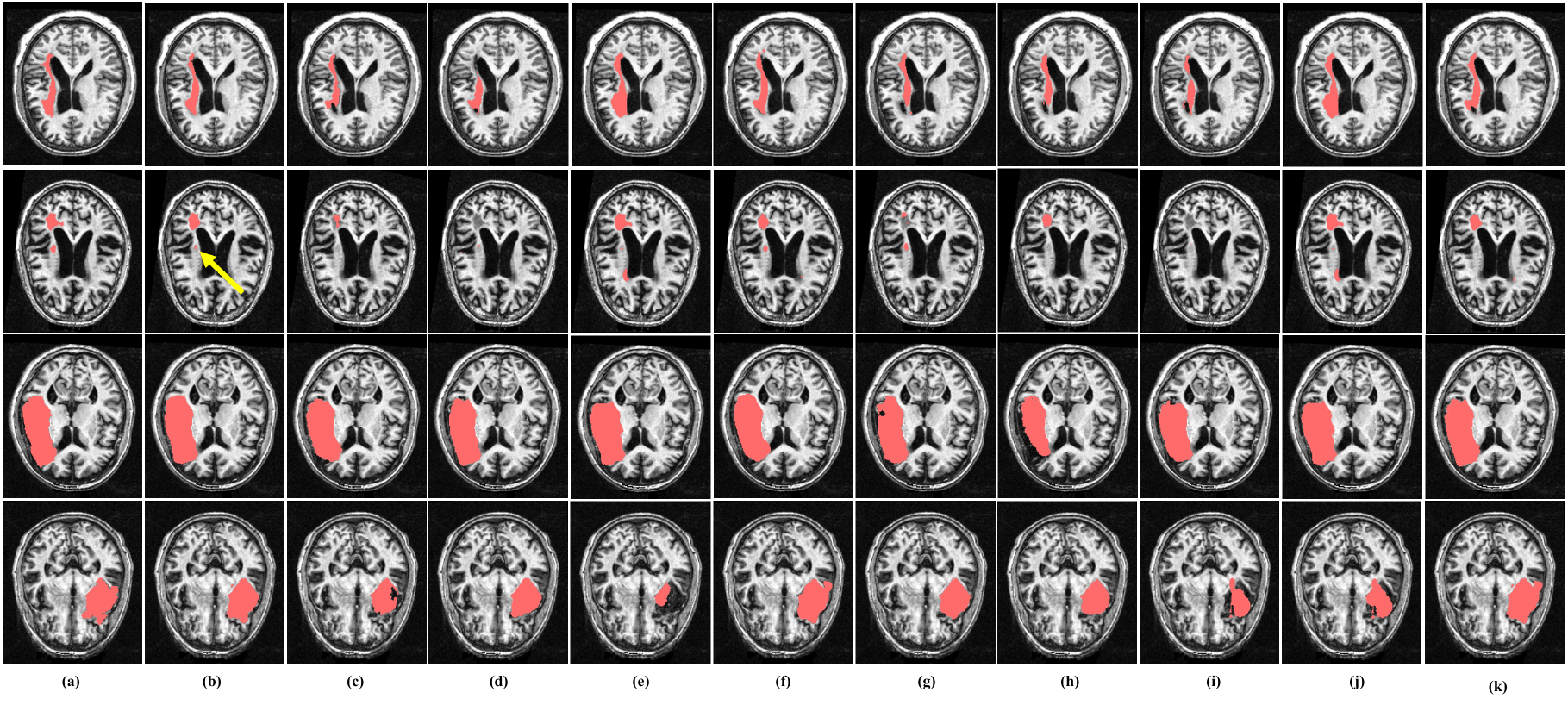}
\caption{Examples of segmentation results on ATLAS dataset. The rows show the performance for cases involving different lesion locations, sizes, or numbers. The first and second columns represent the input images and ground-truth segmentation masks, and the remaining columns represent different segmentation methods: (\textbf{a}): Input MR image with ground-truth segmentation mask; (\textbf{b}): \methodname (Ours); (\textbf{c}) U-net; (\textbf{d}) UNet3+; (\textbf{e}) nnU-Net; (\textbf{f}) X-Net; (\textbf{g}) CLCI-Net; (\textbf{h}) DFENet; (\textbf{i}) FACT; (\textbf{j}) Unlearning; (\textbf{k}) RAM-DSIR.}
\label{fig:quality_comparison}
\end{figure*}

\subsubsection{Qualitative Comparison}

Fig.~\ref{fig:quality_comparison} shows some segmentation results from the proposed and the previous methods. The selected slices show different lesion locations, sizes, or quantities. We set the stroke prediction probability threshold to 0.5 following the convention of semantic segmentation. In this experiment, we take Site 5 as the testing set and the other eight sites as the training set. For instance, the second row shows two small lesions in the image. Our method predicts the presence of two lesions with shapes similar to those in the ground truth, but the other methods can only predict one or none of the lesions (note the tiny lesion indicated by the yellow arrow). According to Fig.~\ref{fig:quality_comparison}, our \methodname provides more generalized and robust performance on unseen sites, regardless of lesion variance.

\subsection{Further Evaluation and Analysis}\label{sec:exp:further_results}

\subsubsection{Generalization to Unseen External Sites}\label{sec:exp:generalization}

\begin{table}[t]
\renewcommand{\arraystretch}{1.0}
\caption{Further evaluation on unseen external sites from ATLAS v2.0. The standard deviation behind $\pm$ is across the nine trained models on ATLAS v1.2. Since all p-values in Wilcoxon tests are less than $5$\%, we do not show them in the table.}
\label{tab:generalization}
\centering
\begin{tabular}{rccc}

\shline
\textbf{Method} &\textbf{Dice} &\textbf{Recall} &\textbf{F1-score} \\

\midrule

\multicolumn{4}{c}{\textbf{Site 40 / Siemens Vision / 45 Subjects}} \\
\midrule
U-net~\cite{2015U} & 0.2470{\scriptsize$\pm$0.0332} & 0.1923{\scriptsize$\pm$0.0431} & 0.2064{\scriptsize$\pm$0.0326}\\

Unlearning~\cite{dinsdale2021deep} & 0.3993{\scriptsize$\pm$0.0141} & 0.3546{\scriptsize$\pm$0.0350} & 0.3762{\scriptsize$\pm$0.0346}\\

RAM-DSIR~\cite{zhou2022generalizable} & 0.4129{\scriptsize$\pm$0.0245} & 0.4035{\scriptsize$\pm$0.0301} & 0.3984{\scriptsize$\pm$0.0248}\\

\methodname (ours) & \textbf{0.4301}{\scriptsize$\pm$0.0243} & \textbf{0.4316}{\scriptsize$\pm$0.0475} & \textbf{0.4301}{\scriptsize$\pm$0.0243}\\

\midrule

\multicolumn{4}{c}{\textbf{Site 48 / Philips Achieva / 23 Subjects}} \\
\midrule
U-net~\cite{2015U} & 0.1272{\scriptsize$\pm$0.0352} & 0.0762{\scriptsize$\pm$0.0124} & 0.1325{\scriptsize$\pm$0.0376}\\

Unlearning~\cite{dinsdale2021deep} & 0.3642{\scriptsize$\pm$0.0252} & 0.3793{\scriptsize$\pm$0.0331} & 0.3672{\scriptsize$\pm$0.0346}\\

RAM-DSIR~\cite{zhou2022generalizable} & 0.3812{\scriptsize$\pm$0.0252} & 0.4148{\scriptsize$\pm$0.0237} & 0.3832{\scriptsize$\pm$0.0325}\\

\methodname (ours) & \textbf{0.3894}{\scriptsize$\pm$0.0225} & \textbf{0.4259}{\scriptsize$\pm$0.1792} & \textbf{0.3921}{\scriptsize$\pm$0.0148}\\

\shline

\end{tabular}
\end{table}

To further evaluate the generalization to unseen sites, we use data of Sites 40 and 48 from ATLAS v2.0~\cite{Liew2021.12.09.21267554} used in ISLES 2022~\cite{hernandez2022isles}. The new version contains more new unseen sites that could serve as external data for further evaluation of generalization. The types of MR scanners for Sites 40 and 48 are Siemens Vision and Philips Achieva, respectively.

We compared our method with the baseline U-net~\cite{2015U} and two most competitive methods (Unlearning~\cite{dinsdale2021deep} and RAM-DSIR~\cite{zhou2022generalizable}), as shown in Table~\ref{tab:generalization}. The results further validate more effectiveness and the generalization of our method overall on new sites regardless of the scanner types. Although the difficulty of segmentation varies in different sites, our method provides consistently better performance than other methods in all the metrics. Besides, the Wilcoxon tests for the further evaluation on unseen external sites demonstrate that all p-values are less than 5\%.

\subsubsection{Analysis on Data Imbalance Across Sites}\label{sec:exp:data_imbalanced}

\begin{table}[t]
\renewcommand{\arraystretch}{1.0}
\caption{The result of our method on nine sites with and without balanced sampling}
\label{tab:balanced}
\centering
\begin{tabular}{rccc}

\shline
\bfseries Sampling &\bfseries Dice &\bfseries Recall &\bfseries F1-score\\

\midrule
Non-balanced & \textbf{0.5711}{\scriptsize$\pm$0.1957} & \textbf{0.5977}{\scriptsize$\pm$0.1580} & \textbf{0.5623}{\scriptsize$\pm$0.1926}\\
Balanced & 0.5690{\scriptsize$\pm$0.1925} & 0.5851{\scriptsize$\pm$0.1793} & 0.5568{\scriptsize$\pm$0.1824}\\
p-value\ & 0.7793 & 0.3120 & 0.6879\\

\shline
\end{tabular}
\end{table}
In light of Table~\ref{tab:site}, the number of patients from each site is not balanced. We perform experiments with balanced sampling according to the inverse ratio of the number of patients from each site. The balanced sampling could make each site equally sampled for training. Table~\ref{tab:balanced} shows the results of balanced sampling and our non-balanced sampling. Regarding the p-values for different evaluation metrics in the Wilcoxon tests, there are no statistically significant differences between with and without balanced sampling. This may be due to the gradient reversal layer for backward propagation: when the model initially has a bias towards a special site, the gradient reversal layer overcomes this bias and makes all sites indistinguishable.

\subsubsection{Performance of Individual Sites}

\begin{figure*}[h]
\centering
\includegraphics[width=1\linewidth]{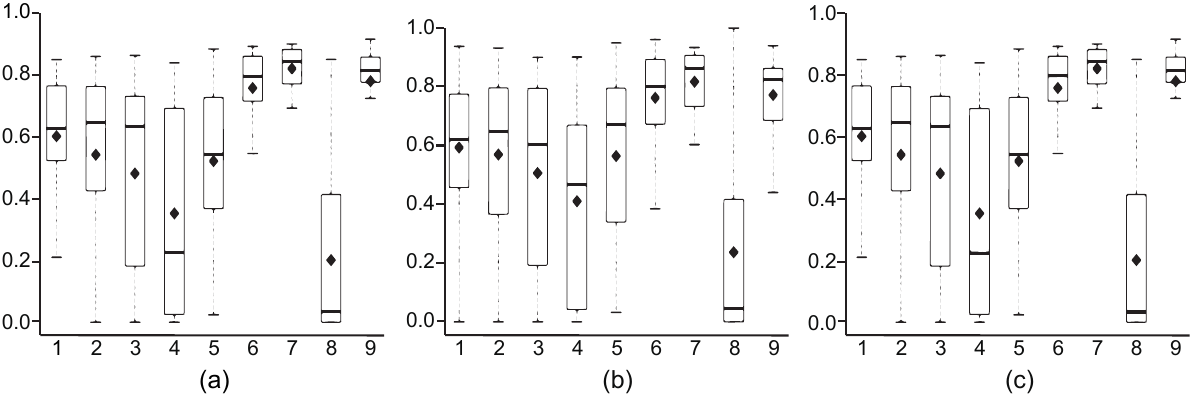}
\caption{The boxplots for the performance of our model on nine sites. (\textbf{a})-(\textbf{c}) represent Dice, recall, and F1-score, respectively. The solid diamond represents the mean.}
\label{fig:boxplot}
\end{figure*}
\begin{figure}[h]
\centering
\includegraphics[width=0.7\linewidth]{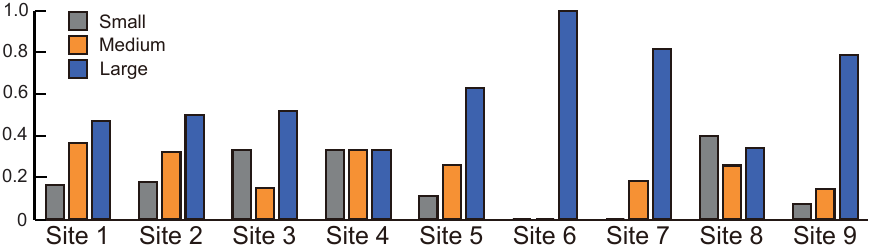}
\caption{The histogram of lesion size in nine sites (the vertical axis represents the percentage). Small lesion: less than 1$\times$10$^3$ voxels; medium lesion: from 1$\times$10$^3$ to 5$\times$10$^3$ voxels; and large lesion: greater than 5$\times$10$^3$ voxels.}
\label{fig:bar}
\end{figure}

Fig.~\ref{fig:boxplot} presents the boxplot of each metric for individual sites. Combined with Table~\ref{tab:site}, the scanner types seem to not directly influence segmentation performance. For instance, the type of MR scanners in Site 1 is the same as Site 8 (GE 750 Discovery), however, there is a significant difference between their segmentation performances. We further analyzed the performance of our model in relation to lesion size. Fig.~\ref{fig:bar} shows the distribution of the lesion size for each site. We can see that the worst performance of the model on Site 8 mainly results from a large proportion of patients with a small lesion. On the other hand, it is easier for the model to achieve a high Dice score for a large lesion. A similar phenomenon was also demonstrated in~\cite{2020Automatic, 2020MI}, though the loss functions, network architecture, and optimization algorithm are well enough selected. For example, Sites 6, 7, and 9 include a large proportion of patients with a large lesion. More importantly, the great variance of lesion size resulted in a large standard deviation across subjects in the metrics.

\subsubsection{Robustness Verification on the Midline Shift}\label{sec:exp:mid}

\begin{figure*}[h]
\centering
\includegraphics[width=0.9\linewidth]{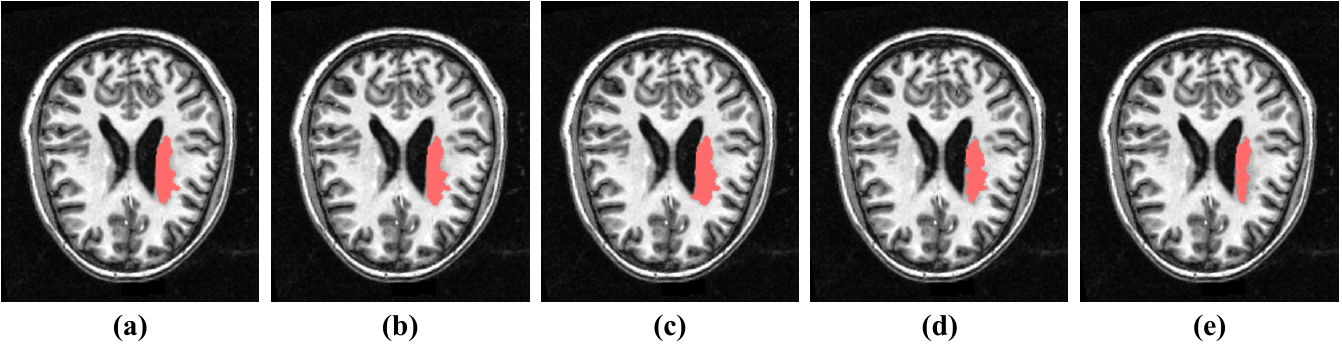}
\caption{Example segmentation results with different rotation angles: (\textbf{a}) Groud truth; (\textbf{b}) Rotation angle is 0°; (\textbf{c}) Rotation angle is 15°; (\textbf{d}) Rotation angle is 30°; (\textbf{e}) Rotation angle is 45°. Note that the rotated images are finally rotated back in reverse for convenient comparison.}
\label{fig:rotation}
\end{figure*}

The released ATLAS dataset has been normalized to standard (MNI-152) space, hence, there would be no severe midline shift in the inputs. To explore a possible situation when the MR image shows a severe midline shift, we further investigate the influence of different clockwise rotation angles around the center of the image on the segmentation performance of SAN-Net, as shown in Table~\ref{tab:midlinle}. Therefore, \methodname with SIDA can deal with a slight shift. When the inputs show a severe midline shift, the Dice coefficient drops from 0.43 to 0.35. Moreover, Fig.~\ref{fig:rotation} presents a set of segmentation results with different rotation angles, which intuitively demonstrates the influence of the rotation angles on the performance of SIDA.

\begin{table}[h]
\renewcommand{\arraystretch}{1.0}
\caption{Comparison among different clockwise rotation angles around the center of the image (the testing site is Site 5).}
\label{tab:midlinle}
\centering
\begin{tabular}{cccc}

\shline
\bfseries Rotation angle ($^{\circ}$) &\bfseries Dice &\bfseries Recall &\bfseries F1-score\\

\midrule
0 & \textbf{0.4396} & \textbf{0.4044} & \textbf{0.4565} \\
15 & 0.4332 & 0.3924 & 0.4518\\
30 & 0.4158 & 0.3873 & 0.4394\\
45 & 0.3572 & 0.3605 & 0.3726\\

\shline
\end{tabular}
\end{table}

\section{Discussion and Conclusion}\label{sec:dis}

In this paper, we propose \methodname for site generalization. First of all, the proposed MAIN can dynamically standardize the input MR images into a site-unrelated style. Then, the classifier with a gradient reversal layer is implemented to guide MAIN and the encoder of U-net to extract site-invariant features. Moreover, a data augmentation method, named SIDA, is embedded within \methodname to help the model locate lesions easily. It is friendly to GPU occupation and computational cost as well. The experimental results on the ATLAS dataset demonstrate that our \methodname shows increased model generalization to unseen sites, and the ablation study validates the effectiveness of each component. 

Here, we acknowledge the following limitations and potential solutions for this work.
\begin{enumerate}
\item Since the preprocessing done by the original authors can reduce site (domain) variance, MAIN assumes the domain shift across multiple sites appears as an affine transformation on intensity distribution. However, the brain MR images from different sites show more complex variations in clinical practice, as shown in Table~\ref{tab:site}. How to extract site-invariant features without preprocessing deserves further investigation.

\item In terms of the robustness of SIDA, there is no patient whose lesion spans the hemispheres so far in the ATLAS dataset. More precisely, as summarized in~\cite{2018A}, the distribution of lesions is 48.4\% in left hemispheres, 43.8\% in right hemispheres, and 7.7\% at other locations such as brainstems or cerebellums. SIDA would be improved based on the involvement of such cases in the training set. 
In addition, as illustrated in Section~\ref{sec:exp:mid}, our \methodname can only deal with a slight midline shift. In practice, a severe midline shift is usually caused by unsuccessful image registration.

\item It is worth noting that the backbone of \methodname is the traditional U-net for a fair comparison. We intend to further design a customized U-net and provide more precise and generalized segmentation, in cooperation with the proposed components.

\item The existing methods mainly conducted experiments on ATLAS v1.2, which has been well-studied. To compare with other methods under the optimal parameter settings, our ablation study and comparison study were mainly based on ATLAS v1.2. We plan to perform more benchmark experiments on ATLAS v2.0, which contains a larger number of stroke patients.

\end{enumerate}

In conclusion, the proposed \methodname  enhances the stroke lesion segmentation performance on MR images from unseen sites, which suggests its potential applicability in clinical practice. In the future, we will be extending current \methodname with federated learning techniques~\cite{dou2021federated, guo2021multi, zerka2021privacy}, which will be a more flexible framework for site generalization without patient privacy.


\end{document}